\documentclass[conference]{IEEEtran}
\IEEEoverridecommandlockouts

\usepackage{cite}
\usepackage{amsmath,amssymb,amsfonts}
\usepackage{algorithmic}
\usepackage{graphicx}
\usepackage{textcomp}
\usepackage{xcolor}
\usepackage{booktabs}
\usepackage{xurl}

\makeatletter

\DeclareRobustCommand*{\RWWauthorrefmark}[1]{\raisebox{0pt}[0pt][0pt]{\textsuperscript{#1}}}%

\makeatother

\def\BibTeX{{\rm B\kern-.05em{\sc i\kern-.025em b}\kern-.08em
		T\kern-.1667em\lower.7ex\hbox{E}\kern-.125emX}}

\usepackage[acronym]{glossaries}
\makenoidxglossaries

\newacronym{cir}{CIR}{Channel Impulse Response}
\newacronym{sir}{SIR}{Signal-to-Interference Ratio}
\newacronym{ieee}{IEEE}{Institute of Electrical and Electronics Engineers}
\newglossaryentry{wifi}{
	name=Wi-Fi,
	description={Wireless Fidelity}
}
\newacronym{cfo}{CFO}{Carrier Frequency Offset}
\newacronym{pbr}{PBR}{Phase-Based Ranging}
\newacronym{los}{LOS}{Line-of-Sight}
\newacronym{nlos}{NLOS}{Non-Line-of-Sight}
\newacronym{awgn}{AWGN}{Additive White Gaussian Noise}
\newacronym{rtt}{RTT}{Round Trip Time}
\newacronym{snr}{SNR}{Signal to Noise Ratio}
\newacronym{drbg}{DRBG}{Deterministic Random Bit Generator}
\newacronym{phy}{PHY}{Physical Layer}
\newacronym{bt}{BT}{Bluetooth}
\newacronym{cs}{CS}{Channel Sounding}
\newacronym{btcs}{BT CS}{Bluetooth Channel Sounding}
\newacronym{ble}{BLE}{Bluetooth Low Energy}
\newacronym{ism}{ISM}{Industrial, Scientific and Medical}
\newacronym{mse}{MSE}{Mean Squared error}
\newacronym{fft}{FFT}{Fast Fourier Transform}
\newacronym{iot}{IoT}{Internet of Things}
\newacronym{btsig}{BT SIG}{Bluetooth Special Interest Group}

\glsunset{btcs}
\glsunset{ism}

\begin{document}
	\title{Channel Modeling for Phase-Based Ranging}

\author{
	\IEEEauthorblockN{
		Till Droemmer\RWWauthorrefmark{1,2}, 
		Markus Gardill\RWWauthorrefmark{1}
	}
	
	\IEEEauthorblockA{
		\RWWauthorrefmark{1}Chair of Electronic Systems and Sensors, 
		Brandenburg University of Technology, Germany\\
		\RWWauthorrefmark{2}Mercedes-Benz AG
	}
}

\maketitle
	
	\begin{abstract}
	This paper presents a Python-based simulation framework for the physical 
	layer 
	of Bluetooth Channel Sounding in accordance with Bluetooth Core 
	Specification 
	v6.2. The simulator implements Mode 3 phase-based ranging across 72 active 
	tones in the 2.4 GHz \gls{ism} band and supports multiple propagation 
	conditions, 
	including free-space, static multipath, and IEEE 802.15.4a stochastic 
	multipath 
	channels. In addition, it models relevant hardware and interference effects 
	such as phase noise, phase ramp, carrier frequency offset, IQ imbalance, 
	ADC 
	quantization, and narrowband interference from co-located Wi-Fi systems, 
	with 
	parameters derived from commercial SoC datasheets and field measurements. 
	The 
	framework enables controlled and repeatable analysis of 
	physical layer effects that are difficult to isolate in over-the-air 
	experiments and are not sufficiently represented in existing higher-level 
	simulation tools. This work focuses exclusively on physical layer signal 
	generation, 
	channel modeling, and impairment injection, providing a foundation for 
	future 
	studies on Bluetooth Channel Sounding ranging and localization algorithms.
	\end{abstract}

	
	\section{Introduction}
	\gls{bt} \gls{cs}, introduced in Bluetooth Core Specification v6.0 
	and refined in Core Specification v6.1 and v6.2, enables sub-meter distance 
	measurement 
	between devices using \gls{pbr} in the 2.4~GHz \gls{ism} band 
	\cite{b1}. 
	Unlike traditional \gls{rtt}-only methods \cite{b2}, 
	\gls{pbr} exploits frequency-domain phase measurements across multiple 
	frequencies to calculate the distance between two devices.
	
	The development and optimization of \gls{btcs} ranging methods require 
	extensive testing across diverse channel conditions, hardware impairments, and 
	interference scenarios. Prototyping based on real-world measurements is 
	time-consuming and expensive, and even slight variations in the test setup 
	influence outcomes, making controlled, repeatable evaluation difficult. 
	While real-hardware studies such as \cite{b3} 
	provide valuable experimental insight and openly available datasets, 
	measurements inherently cover only a limited set of conditions; 
	channel parameters, hardware impairments, and interference levels cannot be 
	varied independently or exhaustively. On the simulation side, the MATLAB
	Bluetooth Toolbox \cite{b4} supports specification-compliant
	\gls{btcs} PHY waveforms covering various multi-frequency channel stepping 
	schemes,
	bidirectional tone exchange, and both \gls{pbr} and
	\gls{rtt} distance estimation. However, it is primarily oriented toward system-level
	simulation and does not focus on the control of individual
	frequency-domain channel and impairment mechanisms, such as multipath, IQ imbalance,
	ADC quantization, and co-located \gls{wifi}/OFDM interference. Fine-grained, independent control of
	physical layer effects is essential for isolating the impact of individual
	impairments on \gls{pbr} ranging performance, which motivates the need for a
	dedicated, high-fidelity simulation framework.

	This paper presents a Python-based \gls{btcs} simulator specifically
	designed for research on \gls{pbr} ranging, scoped exclusively to
	physical layer signal generation, channel modeling, and impairment
	injection. The simulator does not evaluate ranging accuracy or estimator
	performance; rather, it provides the 
	physical layer accurate channel responses that serve as a foundation for 
	future
	ranging and localization studies. The architecture is modular, enabling
	independent variation of channel conditions, and channel impairments.
	More specifically, the simulator includes a specification-compliant
	implementation of the \gls{btcs} Mode~3 \gls{pbr} physical layer together
	with comprehensive impairment modeling derived from characteristics of
	commercial SoCs, including phase noise, \gls{cfo},
	IQ imbalance, and ADC quantization effects. In addition, it incorporates
	a detailed narrowband interference model for \gls{wifi} IEEE~802.11 OFDM systems
	that accounts for both CSMA/CA duty cycling and \gls{cfo} effects. Finally, the 
	simulator is
	qualitatively validated against an open real-hardware dataset, demonstrating
	that it captures the dominant physical effects observed in practical
	\gls{btcs} deployments.
	
	The remainder of this paper is organized as follows. Section~II
	describes the simulation framework and component models. Section~III 
	validates the simulator qualitatively against real-hardware measurements.
	Section~IV concludes.
	
	\section{Simulation Framework}
	\subsection{Overview}
	The simulator takes a scenario configuration as input, specifying the true
	distance, channel model, and impairment parameters, and produces as output
	the complex frequency-domain tone responses that a ranging estimator would
	receive. During a \gls{btcs} Mode~3 procedure, the receiver coherently
	integrates the received unmodulated CW signal over the phase measurement
	period $T_\text{PM}$ for each tone, yielding one complex IQ sample $Y[k]$
	per tone that represents the channel response at frequency $f_k$. The
	simulator operates entirely at this level, modeling the 72-element complex
	vector $\mathbf{Y} \in \mathbb{C}^{72}$, without constructing time-domain
	waveforms or simulating a receiver sample clock.

	The propagation channel $H[k]$ is first generated according to the selected
	channel condition (Section~\ref{sec:channel}). A chain of impairments
	$f_\text{imp}$ is then applied to yield the received tone response:
	\begin{equation}
		\resizebox{0.9\hsize}{!}{$
		Y[k] = \mathrm{Q}_B\!\left[\bigl(g_1 H[k] + g_2 H^*[k]\bigr)
		e^{\,j(\delta_k + \phi_{\text{ramp},k})} + n_k + I[k]\right]
		$}
		\label{eq:toplevel}
	\end{equation}
	where $n_k$ is \gls{awgn}, $\delta_k$ is phase noise, $\phi_{\text{ramp},k}$
	is the \gls{cfo}-induced phase ramp, $g_1,g_2$ are IQ imbalance gain
	factors, $I[k]$ is narrowband interference, and $\mathrm{Q}_B[\cdot]$
	denotes per-component uniform quantization to $B$~bits. Each impairment is
	described in Section~\ref{sec:impairments}; all or any subset may be
	enabled independently.

	Mode~3 simulates a bidirectional tone exchange: the impairment chain is
	applied twice with independent noise realizations, producing $Y_A[k]$ 
	(Initiator
	$\to$ Reflector) and $Y_B[k]$ (Reflector $\to$ Initiator) over the same
	channel realization $H[k]$. These are combined per specification as
	$Y_\text{comb}[k] = Y_A[k] \cdot Y_B[k]$, which cancels the
	symmetric phase contributions of both devices and doubles the
	range-dependent phase slope. Specification-compliant timing parameters
	for v6.2 are used throughout: $T_\text{PM}$ determines the noise
	bandwidth and hence the per-tone \gls{snr}, while $T_\text{step}$ governs
	how much the \gls{cfo}-induced phase ramp $\phi_{\text{ramp},k}$
	accumulates between consecutive tone measurements.
	\begin{figure}[t]
		\centerline{\includegraphics[width=0.49\textwidth]{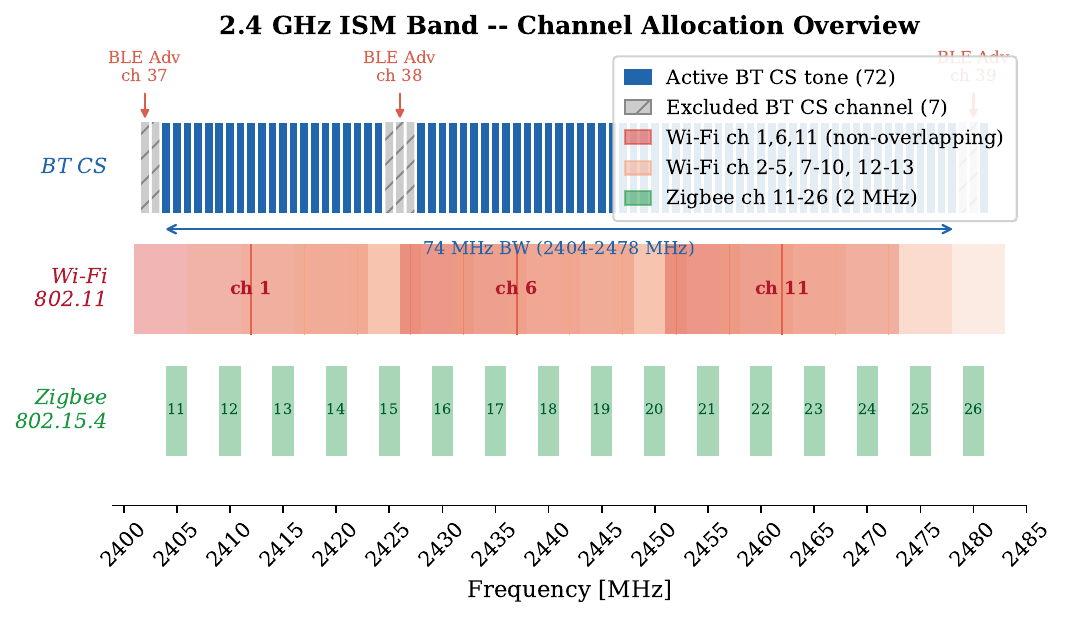}}
		\vspace{-0.31cm}
		\caption{Bluetooth Channel Sounding tone grid with excluded channels to 
			protect the \gls{ble} advertising channels, and \gls{wifi} and 
			Zigbee 
			channels that share the 2.4~GHz \gls{ism} band.}
		\label{fig:channelspectrum}
		\vspace{-0.31cm}
	\end{figure}
	\subsection{Tone Grid Definition}
	In the Bluetooth Core Specification, \gls{btcs} is specified to operate 
	on 79 potential channels spanning 2402--2480~MHz with 
	1~MHz 
	spacing, depicted in Fig.~\ref{fig:channelspectrum}. Seven channels are 
	excluded per specification, \gls{btcs} channel 
	indices 0, 1, 23, 
	24, 25, 77, 78, to prevent interference on the \gls{ble} advertising 
	channels at the same frequencies \cite{b5}.
	Per specification, tones are exchanged in pseudorandom order to decorrelate
	the systematic phase ramp from frequency index.
	The simulator reproduces this pseudorandom tone ordering, ensuring that the
	phase ramp accumulates with respect to measurement time rather than
	frequency index.
	
	\subsection{Channel Conditions}
	\label{sec:channel}
	\begin{figure}[t]
		\centerline{\includegraphics[width=0.49\textwidth]{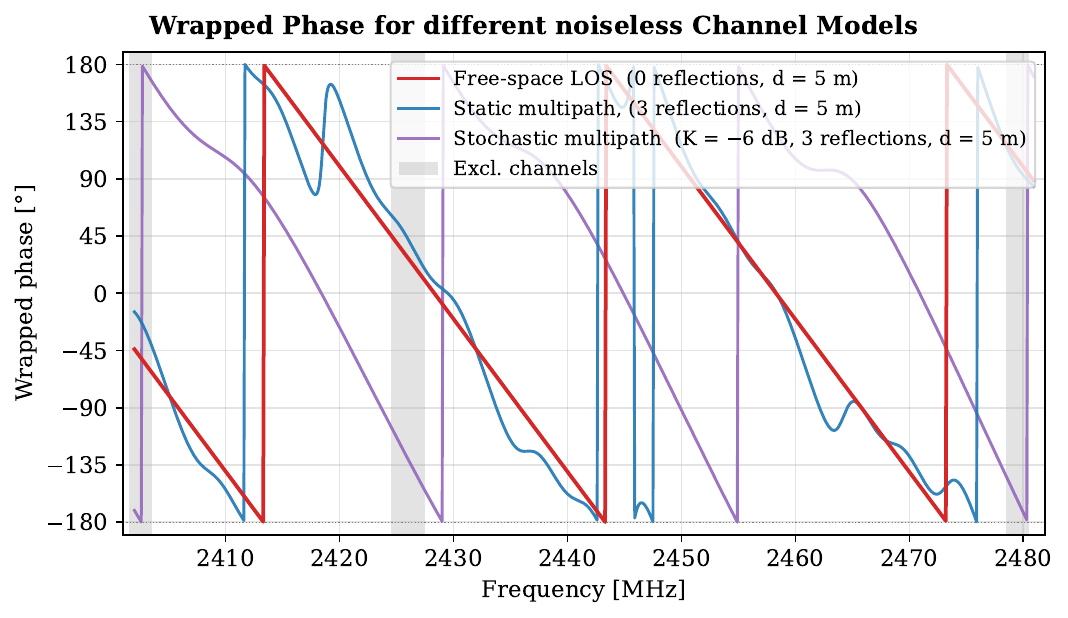}}
		\vspace{-0.31cm}
		\caption{Wrapped phase vs.\ frequency for the three channel conditions:
		free-space, static multipath, and stochastic multipath.}
		\label{fig:channelconditions}
		\vspace{-0.31cm}
	\end{figure}
	To cover the range of environments encountered in \gls{btcs} deployments,
	the simulator provides three selectable channel conditions: 
	a free-space baseline, a deterministic multipath model, and a stochastic multipath model.
	
	\textbf{Free-Space:} The channel is modeled as a single unobstructed
	propagation path with unity amplitude. The phase at each tone is defined by the
	two-way propagation delay $\tau = 2d/c$ via $H(f_k) = e^{-j2\pi f_k \tau}$. This
	causes a phase that increases linearly with frequency and wraps at
	$\pm 180^\circ$, producing the wrapped linear phase ramp visible as the
	red curve in Fig.~\ref{fig:channelconditions}.
		
	\textbf{Static Multipath:} User-defined taps are specified as tuples of 
	(delay~[ns], 
	amplitude, phase~[deg]), where the amplitude is given on a linear scale. 
	The frequency response is computed as
	\begin{equation}
		H(f_k) = \sum_{p=1}^{N_p} a_p e^{j(\phi_p - 2\pi f_k \tau_p)}
	\end{equation}
	where $a_p$, $\phi_p$, $\tau_p$ are amplitude, phase offset, and delay of 
	path $p$.
	In Fig.~\ref{fig:channelconditions}, the static multipath channel with one 
	\gls{los} path and three \gls{nlos} paths is shown as the blue curve. It 
	clearly shows the impact of multipath propagation on the phase of 
	the signal, leading to altered phase slope and additional phase wraps 
	caused by phase deviations close to $\pm 180^\circ$.
	
	\textbf{Stochastic Multipath:} Implements the IEEE~802.15.4a CM1 indoor
	residential channel model \cite{b6}.
	Scattered path amplitudes are drawn from a complex 
	Gaussian distribution; the overall channel is Rician, 
	with a deterministic \gls{los} component weighted by the K-factor.
	In Fig.~\ref{fig:channelconditions}, this channel condition is shown by 
	the 
	purple curve. When comparing its slope to the static multipath, the 
	stochastic multipath channel exhibits a smoother phase response whose slope more closely resembles the
	free-space condition. This behavior arises from the shorter excess delays 
	of the reflection 
	paths in the stochastic multipath condition, as it is a residential model.
	
	\subsection{Channel Impairments}
	\label{sec:impairments}
	\begin{figure}[!h]
		\centerline{\includegraphics[width=0.49\textwidth]{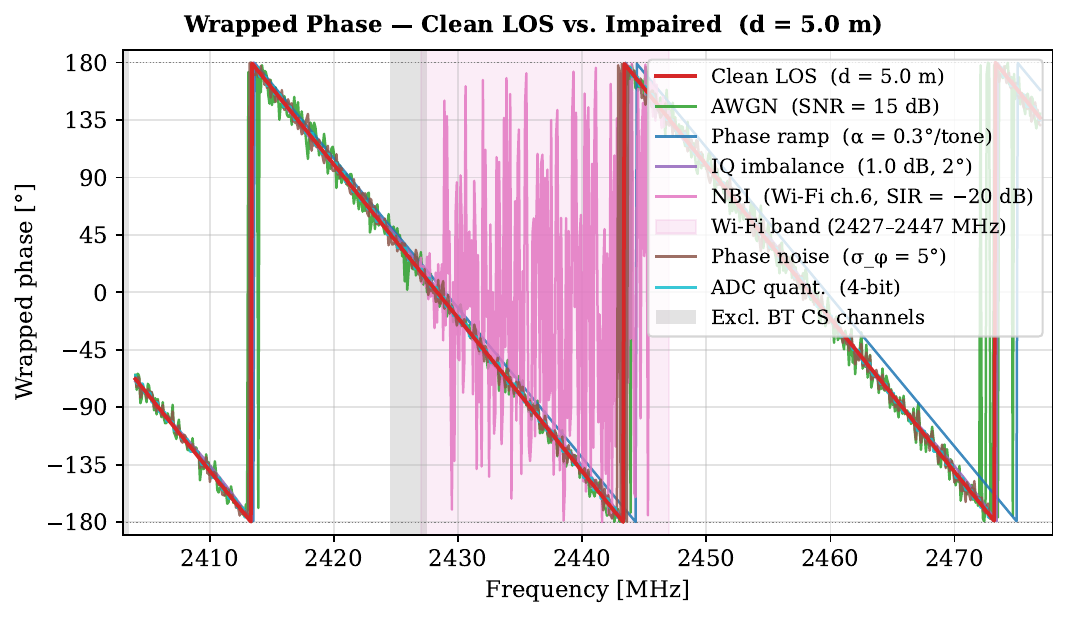}}
		\vspace{-0.31cm}
		\caption{Influence of the different channel impairments on the 
		frequency response of a signal.}
		\label{fig:channelimpairments}
		\vspace{-0.31cm}
	\end{figure}
	
	The impairments in~\eqref{eq:toplevel} can be enabled individually or in
	combination. Fig.~\ref{fig:channelimpairments} shows the effect of each
	impairment on the wrapped phase of a clean free-space \gls{los} channel,
	isolating each distortion relative to the unimpaired reference.

	\textbf{\gls{awgn}:} Thermal noise at the receiver adds an independent
	complex Gaussian term to each tone, giving $Y[k] = H[k] + n_k$.
	Scaled to achieve a target per-tone \gls{snr}, for a received signal
	power $P_s$, the noise at tone $k$ is
	\begin{equation}
		n_k \sim \mathcal{CN}\!\left(0,\, \frac{P_s}{\mathrm{SNR_{lin}}}\right).
	\end{equation}
	The resulting random phase scatter is visible across all 
	tones, as shown by the green curve in Fig.~\ref{fig:channelimpairments}.

	\textbf{Phase Noise:} Oscillator and phase-locked loop imperfections
	perturb the phase of each tone independently, contributing the $e^{j\delta_k}$
	term in~\eqref{eq:toplevel}. This gives $Y[k] = H[k]\cdot e^{j\delta_k}$
	when acting alone with:
	\begin{equation}
		\phi_{\text{meas},k} = \phi_{\text{true},k} + \delta_k, \quad 
		\delta_k \sim \mathcal{N}(0,\, \sigma_\phi^2)
	\end{equation}
	where $\phi_{\text{true},k}$ is the ideal phase of tone $k$ as determined 
	by the channel, $\phi_{\text{meas},k}$ is the phase after perturbation, 
	and $\delta_k$ is an independent zero-mean Gaussian perturbation with 
	standard deviation $\sigma_\phi$.
	The standard deviation $\sigma_\phi$ represents the integrated phase noise 
	power over the $T_\text{PM}$ measurement interval. Lab-grade oscillators achieve 
	$\sigma_\phi \approx 1^\circ$; consumer SoCs under thermal stress can reach 
	$\sigma_\phi \approx 5$--$8^\circ$. Unlike \gls{awgn}, 
	phase noise perturbs only the phase of each tone while leaving its amplitude unchanged. 
	As the phase noise is lower than the \gls{awgn} level, it is barely visible in 
	Fig.~\ref{fig:channelimpairments} as irregular phase deviations.
	
	\textbf{Phase Ramp:} A residual \gls{cfo}
	between the initiator and reflector crystals, combined with VCO drift,
	contributes the $e^{j\phi_{\text{ramp},k}}$ term in~\eqref{eq:toplevel},
	producing a linear phase ramp across the tone grid:
	\begin{equation}
		\phi_{\text{ramp},k} = 2\pi\,\Delta f_\text{CFO}\cdot k \cdot T_\text{step}
	\end{equation}
	where $k = 0, 1, \ldots, N-1$ is the measurement index,
	$\Delta f_\text{CFO}$ is the net \gls{cfo} in Hz,
	representing the combined contribution of crystal mismatch and VCO drift,
	drawn from $\mathcal{N}(0,\,\sigma_{\Delta f}^2)$. The corresponding per-step
	phase increment in degrees is $\sigma_\text{ramp} = 360^\circ \cdot
	\sigma_{\Delta f} \cdot T_\text{step}$, yielding
	$\sigma_\text{ramp} = 0.05$--$0.7^\circ$/step for typical consumer
	hardware. As seen in 
	Fig.~\ref{fig:channelimpairments}, the phase ramp manifests as a constant 
	additional tilt added to the channel's phase slope, shown by the 
	blue curve.
	
	\textbf{IQ Imbalance:} Amplitude and phase mismatch between the I and Q
	branches of the receiver front-end introduce the $g_1, g_2$ terms
	in~\eqref{eq:toplevel}, distorting each tone's complex value as
	\begin{equation}
		Y_\text{out} = g_1 Y + g_2 Y^*
	\end{equation}
	where $g_1$ and $g_2$ are complex gain factors determined by the amplitude 
	imbalance $\varepsilon$ and phase mismatch $\varphi_\text{IQ}$. For typical 
	consumer chips (amplitude imbalance: 0.5--1.0~dB, phase imbalance: 
	1.5--2.0$^\circ$ \cite{b7}), the conjugate image 
	term $g_2 Y^*$ remains well below the 
	thermal noise floor at moderate \gls{snr}, but becomes visible at high \gls{snr} 
	as a small, frequency-dependent phase distortion. As the IQ imbalance is small 
	compared to the thermal noise, its effect is barely noticeable, as illustrated in 
	Fig.~\ref{fig:channelimpairments}.
	
	\textbf{ADC Quantization:} This corresponds to the $\mathrm{Q}_B[\cdot]$
	operator in~\eqref{eq:toplevel}. For each tone, the hardware reports one
	complex IQ measurement: the in-phase and quadrature components of $Y[k]$,
	which are each quantized independently to $B$~bits using uniform
	mid-tread quantization over a fixed full-scale range. This is a
	frequency-domain operation: no time-domain waveform or receiver sample
	clock is simulated. The quantization error per component is bounded by
	$\pm\Delta/2$, where $\Delta = 2^{1-B}$ is the least significant bit step. Typical \gls{btcs} SoCs use 12--16 bits 
	\cite{b8,b9,b10}, for which 
	the quantization noise is negligible compared to thermal noise. Coarser 
	quantization (e.g., 4--6 bits) produces a visible staircase distortion in the 
	phase response, shown in Fig.~\ref{fig:channelimpairments}.
	
	\textbf{Narrowband Interference:} Co-located \gls{ism} band transmitters 
	such
	as \gls{wifi} and Zigbee contribute the additive interference term $I[k]$
	in~\eqref{eq:toplevel}. A co-located \gls{wifi} 802.11n/ac access point is
	modeled via its full OFDM subcarrier structure: each 20~MHz channel comprises 64 subcarriers with 
	312.5~kHz spacing. The interference contribution to overlapping \gls{btcs} 
	tones is
	\begin{equation}
		I(f_k) = P_{\text{int}} \sum_{i=-32}^{31} \mathrm{sinc}\!\left(\frac{f_k - f_i}{\Delta f_\text{sub}}\right)
		e^{j(\phi_i + 2\pi \Delta f_{\text{CFO}}\, t)}
	\end{equation}
	where $\Delta f_\text{sub} = 312.5$~kHz is the OFDM subcarrier spacing,
	$\Delta f_\text{CFO}$ is drawn uniformly from $\pm20$~ppm crystal tolerance, and
	$\phi_i$ are random OFDM symbol phases. The intermittent activity of the 
	access point is captured via a Bernoulli process with configurable duty cycle 
	(default: 0.8 for a busy AP). In Fig.~\ref{fig:channelimpairments}, 
	narrowband interference appears as large, localized phase outliers on the 
	subset of \gls{btcs} tones that overlap with the \gls{wifi} channel. It
	is shown by the pink curve, denoted as NBI for \gls{wifi} channel 6 between 
	2427 and 2447~MHz.

\section{Evaluation}

	\subsection{Qualitative Validation Against Real Hardware}
	\label{sec:validation}
	Before applying the simulator systematically, its physical realism is 
	assessed
	by comparing simulated phase responses against real \gls{btcs} measurements.
	The open dataset published by 
	Wieme~\cite{b3}
	provides raw IQ measurements captured with Nordic nRF54L15-DK devices in 
	different environments, including
	a 300~m² warehouse-like environment \cite{b11}. The measurements were taken 
	with the devices operating in Mode~3, using the full 72-tone
	sequence.
		\begin{figure}[t]
			\centerline{\includegraphics[width=0.49\textwidth]{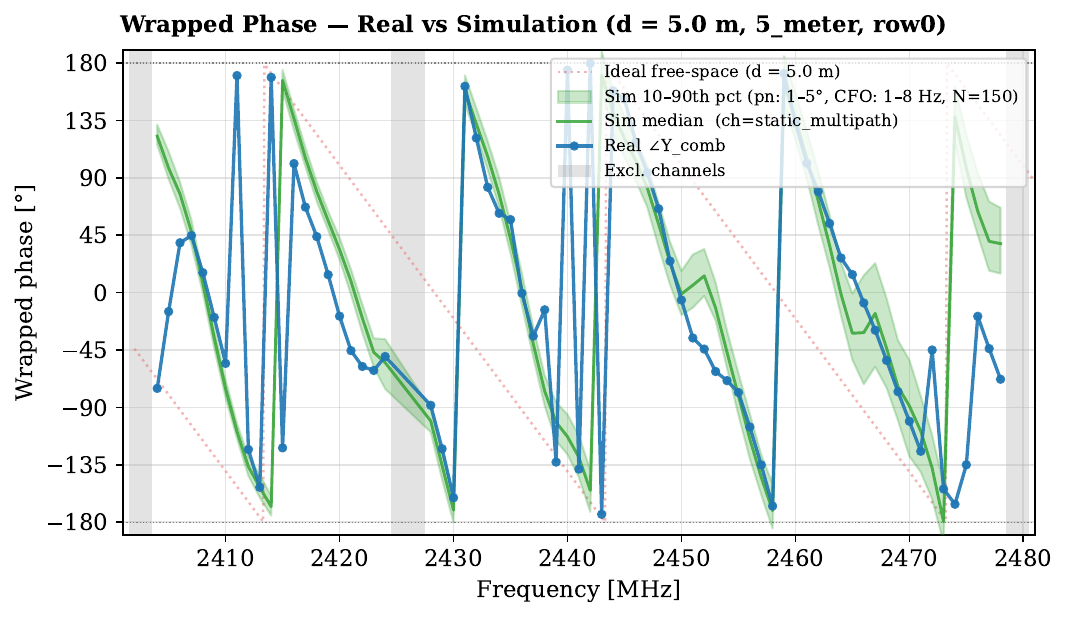}}
			\vspace{-0.31cm}
			\caption{Wrapped phase vs.\ frequency: single real measurement from 
			the
			Wieme dataset (5~m \gls{los}, nRF54L15-DK) overlaid with a
			representative simulated realization using a matched static 
			multipath
			configuration.}
			\label{fig:realvssim}
			\vspace{-0.31cm}
		\end{figure}

	Fig.~\ref{fig:realvssim} shows the ideal free-space reference, a single real
	measurement from the Wieme dataset, and one simulated realization using a
	matched static multipath configuration.
	The reflection paths of the simulation are derived from the warehouse 
	geometry, 
	a dense warehouse
	with metal racks, ceiling infrastructure, and reflective walls 
	\cite{b11}. The paths include a
	dominant shelf reflection, weaker side-wall and end-wall returns, and 
	long
	reverberant tails. 
	
	The nRF54L15-DK hardware parameters are approximated as 
	follows: 20~dBm TX power
	including the nRF21540 front-end module. As the real values are not known, 
	the 
	ADC values are varied between 12 and 16 bits, the \gls{cfo} values are 
	varied 
	between 1 and 8~Hz, and the phase noise is varied between $1^\circ$ and 
	$5^\circ$. To reduce the influence of outlying values, only the 
	10th-to-90th-percentile confidence band of the simulation is plotted in 
	Fig.~\ref{fig:realvssim}.	
	The ideal free-space reference shows a smooth, regularly wrapping phase ramp
	across the 72~active tones.
	Both the real and simulated curves depart significantly from this reference,
	exhibiting a steeper overall phase slope, a larger number of phase wraps, 
	and
	large rapid phase excursions between adjacent tones.
	The real and simulated curves do not follow each other on a tone-by-tone
	basis, but share the same qualitative character: irregular, heavily 
	distorted
	phase profiles that are clearly distinct from the clean free-space ramp.
	Both curves indicate a steeper slope, and therefore a larger apparent 
	distance
	than the true 5~m, which is a common effect of multipath propagation
	\cite{b12,b13}.
	
	The differences between the simulation and the real data are expected, as 
	the exact geometry and reflectivity of every surface in
	the warehouse are unknown, so the assumed simulation model cannot reproduce 
	the
	specific phase pattern of the real channel realization.
	The purpose of this comparison is not a point-by-point match but a 
	validation
	of qualitative behavior. Both curves exhibit the same type and degree of 
	phase
	distortion, confirming that the simulator is capable of producing channel
	responses with the characteristic morphology of a real dense-multipath
	\gls{btcs} deployment.

\section{Conclusion}
This paper presented a Python-based simulation framework for the physical layer 
of \gls{btcs}, covering specification-compliant Mode~3
\gls{pbr} operation across all 72 active tones in the 2.4~GHz \gls{ism}
band. The simulator reproduces the dominant physical effects present in real
\gls{btcs} deployments: multipath propagation, hardware impairments, and
narrowband interference within a controlled, reproducible environment that
is difficult to achieve with over-the-air measurements alone. Qualitative
validation against the open nRF54L15-DK dataset confirmed that the channel
and impairment models capture the essential phase morphology of real
hardware.

The framework is intentionally scoped to the physical layer, producing the
complex frequency-domain channel responses used by ranging estimators
without itself evaluating ranging accuracy. This separation makes the
simulator a suitable basis for future work. The most immediate extension
is to systematically benchmark ranging estimators across various scenarios.
A second direction is tighter coupling with device energy models
to enable joint optimization of measurement duration and ranging
accuracy, which is relevant for battery-constrained Internet of Things nodes.

\end{document}